\documentclass[12pt]{article}

\usepackage{graphicx,psfrag,epsf,color}
\usepackage{amsmath,amssymb,amsfonts}
\usepackage{array}
\usepackage{cite}
\usepackage{multirow}
\usepackage{rotating}

\newcounter{MBQ}

\newcommand{\be}{\begin{equation}}
\newcommand{\ee}{\end{equation}}
\newcommand{\bea}{\begin{eqnarray}}
\newcommand{\eea}{\end{eqnarray}}
\newcommand{\bi}{\begin{itemize}}
\newcommand{\ei}{\end{itemize}}
\newcommand{\ben}{\begin{enumerate}}
\newcommand{\een}{\end{enumerate}}
\newcommand{\bt}{\begin{tabular}}
\newcommand{\et}{\end{tabular}}

\newcommand{\sgl}{\tilde{g}}
\newcommand{\sq}{\tilde{q}}
\newcommand{\medge}{M_{\rm edge}}
\newcommand{\msgl}{m_{\tilde{g}}}
\newcommand{\msq}{m_{\tilde{q}}}
\newcommand{\gasq}{\Gamma_{\sq}}
\newcommand{\A}{\sgl}
\newcommand{\X}{\bar q}
\newcommand{\Y}{q}
\newcommand{\R}{\sq}

\newcommand{\neutralino}{\tilde{\chi}}
\newcommand{\chiindex}{\chi}
\newcommand{\GammaR}{\gasq}
\newcommand{\MR}{\msq}

\begin{document}
\allowdisplaybreaks

\begin{titlepage}

\begin{flushright}
{\small
TUM-HEP-1071/16\\
TTK-16-50\\ 
Cavendish-HEP-16/18\\
24 November 2016
}
\end{flushright}

\vskip1cm
\begin{center}
{\Large \bf Radiative distortion of kinematic 
edges\\[0.1cm] in cascade decays}\\[0.2cm]
\end{center}

\vspace{0.5cm}
\begin{center}
{\sc M.~Beneke$^{a}$, L. Jenniches$^{b}$, A.~M\"uck$^{c}$,} \\ 
and  {\sc M. Ubiali$^{d}$}\\[6mm]
{\it ${}^a$Physik Department T31,\\
James-Franck-Stra\ss e~1, 
Technische Universit\"at M\"unchen,\\
D--85748 Garching, Germany\\
\vspace{0.3cm}
${}^b$
Institut f\"ur Theoretische Physik und Astrophysik, \\
Universit\"at W\"urzburg,
D-97074 W\"urzburg, Germany\\
\vspace{0.3cm}
${}^c$
Institut f\"ur Theoretische Teilchenphysik und Kosmologie,\\
RWTH Aachen University, D-52056 Aachen, Germany\\
\vspace{0.3cm}
${}^d$Cavendish Laboratory, HEP group,\\
University of Cambridge, J.J. Thomson Avenue,\\
Cambridge CB3 0HE, United Kingdom}
\\[0.3cm]
\end{center}

\vspace{0.6cm}
\begin{abstract}
\vskip0.2cm\noindent
Kinematic edges of cascade decays of new particles produced in high-energy 
collisions may provide important constraints on the involved 
particles' masses. For the exemplary case of gluino decay 
$\tilde{g}\to q\bar q \tilde{\chi}$ into a pair of quarks and a neutralino
through a squark resonance, we study the hadronic invariant mass distribution 
in the vicinity of the kinematic edge. We perform a next-to-leading order 
calculation in the strong coupling $\alpha_s$ and the
ratio of squark width and squark mass $\gasq/\msq$, 
based on a systematic expansion in $\gasq/\msq$. The separation 
into hard, collinear and soft contributions elucidates the process-dependent 
and universal features of distributions in the edge region, 
represented by on-shell decay matrix 
elements, universal jet functions and a soft function that depends 
on the resonance propagator and soft Wilson lines.
\end{abstract}
\end{titlepage}

\section{Introduction}
\label{sec:introduction}

\begin{figure}[t]
  \centering
\includegraphics[width=.3\textwidth]{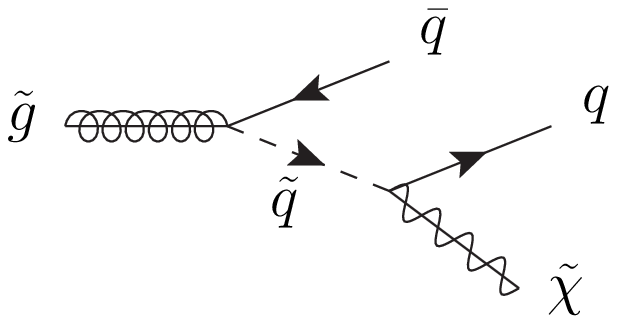}\hspace{2cm}
  \includegraphics[width=.3\textwidth]{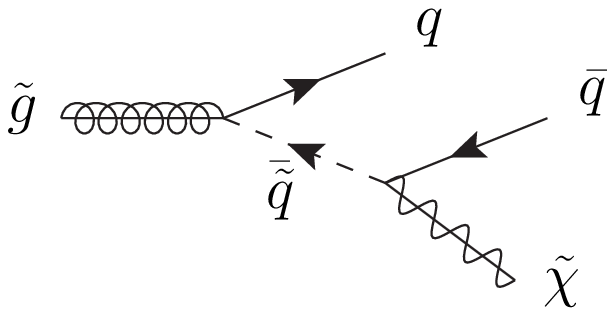}
\caption{Tree diagrams representing the gluino decay 
$\tilde{g}\to q\bar q \neutralino$ through an intermediate squark 
or antisquark resonance.
\label{fig:cascade}}
\end{figure}

The kinematics of particle decay leads to sharp edges in certain 
distributions, whenever the decay proceeds through another intermediate 
resonance. Well-known examples are the invariant mass of the 
lepton pair in squark decay $\tilde{q}\to q \ell^+\ell^-\neutralino$ 
through a neutralino and a slepton 
resonance \cite{Hinchliffe:1996iu,Allanach:2000kt}, 
and the hadronic invariant mass distribution in gluino decay 
$\tilde{g}\to q\bar q \neutralino$ through a squark resonance 
(see diagrams in Figure~\ref{fig:cascade}). 
The latter displays an edge at 
\begin{equation}
\medge^2 = \frac{(\msgl^2-\msq^2)(\msq^2-m_\chiindex^2)}{\msq^2}\,\,. 
\label{edgevalue}
\end{equation}
The sharp feature provides a constraint on the supersymmetric particle 
masses involved in the decay. In practice, the edge will 
be smeared out by detector effects, the extent of which depends 
on the experimental set-up. However, even on purely theoretical 
grounds, the sharp edge is expected to be smoothed by radiative 
corrections and by the width of the intermediate resonance.

In order to predict the spectra locally near the kinematic edge the 
narrow-width approximation for the intermediate resonance cannot 
be applied. This is evident from the fact that the leading radiative 
correction contains a logarithmic singularity 
\begin{equation}
\label{eq:bulktaillogs}
\frac{\alpha_s}{\pi}\ln^2\frac{|M_h^2-M_{\rm edge}^2|}{M_{\rm edge}^2}\,.
\end{equation}
In the edge region, the distribution is sensitive to the resonance 
width $\GammaR$ even when $\GammaR/\MR \ll 1$ and contains 
potentially large logarithms $\ln \MR/\GammaR$.\footnote{The 
width of the resonance determines the extent of the edge region, 
see the following section, which decreases when the resonance is 
longer-lived. Incidentally, the singular logarithms were not 
observed in the next-to-leading order QCD calculation of the process 
$\tilde{q}\to q \ell^+\ell^-\neutralino$ in the 
narrow-width approximation \cite{Horsky:2008yi}, since the 
distribution was binned in bin sizes larger than the width.} A 
reliable theoretical framework must account for the presence of the scale 
$\GammaR$. Radiation and interference effects lead to a 
distortion of the distribution near the kinematic edge.

In this work we quantify this distortion. We define the edge region and 
study the factorization property of the hadronic invariant mass
distribution at leading order in the expansion in the ratio 
$\GammaR/\MR$. The distribution is then computed at next-to-leading 
order (NLO) in the strong coupling $\alpha_s$ and leading order (LO) in 
$\GammaR/\MR$, and at NLO in $\GammaR/\MR$ but LO in $\alpha_s$. 
The resummation of logarithms of $\GammaR/\MR$ is left to future 
work. Our result therefore applies when  $\GammaR/\MR$ is small, but not 
extremely small. We plan to present further details and  
results in a longer technical write-up~\cite{longpaper}. 


\section{The edge region}
\label{sec:fluxes}

\subsection{Kinematics}
\label{sec:kinematics}

\begin{figure}[t]
  \centering
  \includegraphics[width=.6\textwidth]{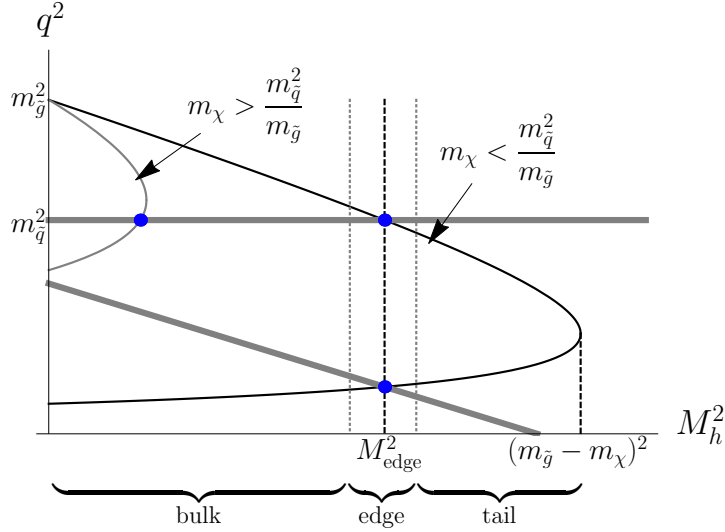}
\caption{Tree-level phase space in the squark momentum squared $q^2$ and 
the hadronic invariant mass squared $M_h^2$. For the case 
$m_{\chiindex}<\msq^2/m_{\tilde g}$ we show the bulk, edge and 
tail regions and the lines (in thick grey) where the (anti)squark is 
on-shell. The intersection of these lines with the phase-space 
boundary (black dots) defines the edge invariant mass, which is shown 
also for the case $m_{\chiindex}>\msq^2/m_{\tilde g}$. 
The non-horizontal grey line corresponds (for the case 
$m_{\chiindex}<\msq^2/m_{\tilde g}$ only) to the resonant value of 
the second diagram in Figure~\ref{fig:cascade}.
\label{fig:phasespace}}
\end{figure}

We consider the gluino decay chain $\tilde{g}\to\sq \,(\to q+\neutralino)+
\bar q$ through an intermediate squark resonance into a  
neutralino. At tree level, the neutralino is accompanied by a quark-antiquark 
pair with invariant mass $M_h$. The edge value~(\ref{edgevalue}) is the 
maximal value the hadronic invariant mass can take for tree kinematics, 
when the squark momentum $q$ is on-shell, $q^2=\msq^2$. The 
tree-level phase space is shown in Figure~\ref{fig:phasespace}. The edge 
value naturally divides the hadronic invariant mass into three regions. The 
``edge region'' is the strip of width ${\cal O}(\gasq)$, in 
which the squark propagator can remain resonant, but the distribution is 
sensitive to the precise virtuality of the propagator. Values $M_h>\medge$ up 
to $\msgl-m_{\chiindex}$ are accessible only when the squark propagator is 
off-shell. We refer to this as the ``tail 
region''. The tree-level distribution falls off rapidly in this region. 
Finally, the region of hadronic invariant mass below the edge region 
is called the ``bulk region''. In this region the shape of the bulk of 
the invariant mass distribution is determined by the cascade of two decay 
processes through an intermediate on-shell squark. 

The edge region is the only region that requires a special treatment, because 
it is intrinsically sensitive to the scale of the resonance width, which 
enters the resonance propagator. While in the bulk region the resonance is 
also on-shell, the propagator can still be expanded in the distribution 
sense, treating $\gasq$ as small. At leading order in $\gasq/\msq$, 
this amounts to the narrow-width approximation. In the tail region, on the 
other hand, the distribution is power-suppressed.

At tree level, invariant masses in the edge region can be produced 
in two ways. For resonant squarks the edge value is attained if the quark 
and antiquark are back-to-back, since the invariant mass increases with 
the angle $\theta$ between the quark and antiquark momenta. Alternatively, 
$\medge$ can also be achieved by $q^2>\msq^2$ or $q^2<\msq^2$, in which 
case $\cos\theta$ does not need to be $-1$. However, this 
contribution is power-suppressed due to the off-shell squark propagator. 
Whether $q^2$ must be larger or smaller than $\msq^2$ depends on whether the 
neutralino mass is larger or smaller than $\msq^2/\msgl$. The value of the 
neutralino mass also determines the resonant decay kinematics in the 
edge region. For small $m_{\chiindex}<\msq^2/\msgl$, the neutralino momentum 
is aligned with the antiquark momentum, otherwise with the quark momentum.

Since the gluino and neutralino are Majorana fermions, there is another 
decay chain, $\sgl\to\bar{\sq} \,(\to \bar q+\neutralino)+q$, where the 
quark and antiquark momenta are interchanged and the resonance is an 
antisquark (see second diagram in Figure~\ref{fig:cascade}), which interferes 
with the squark resonance chain. At LO in $\gasq/\msq$, however, the two 
processes can be treated as independent and contribute the same amount. The 
reason for this is that the interference of the two amplitudes necessarily 
requires one of the squark propagators to be off-shell, and hence is 
$\gasq/\msq$ suppressed. We therefore focus on the first decay chain.

\subsection{Factorization and leading regions}
\label{sec:factorization}

When the squark width is set to zero the invariant mass distribution 
drops to zero discontinuously at the edge value, which is unphysical. Our aim 
is to describe the shape of this distribution correctly at leading order in 
$\gasq/\msq$, including radiative corrections. 

We already noted that the quark and antiquark must be nearly back-to-back 
at tree level. It is evident that tree-level kinematics is not changed, 
if a) the gluino and squark decay vertices are modified by hard-virtual 
corrections, b) the quark and antiquark develop into jets by collinear 
emissions, and c) soft gluons connect all strongly interacting particles 
in the squared amplitude. We therefore introduce the hard $(1,1,1)$, 
collinear $(1,\lambda,\sqrt{\lambda})$, 
anti-collinear $(\lambda,1,\sqrt{\lambda})$ 
and soft $(\lambda,\lambda,\lambda)$ regions, where 
$\lambda=\gasq/\msq$.\footnote{We do not distinguish $\msgl$ and 
$\msq$ for the purpose of power counting.} Here, following soft-collinear 
effective theory (SCET) notation \cite{Bauer:2000yr,Beneke:2002ph}, 
we introduced two light-like vectors, $n_{\pm}^2=0$, and decomposed a 
four-vector into components $(n_+p,n_-p, p_\perp)$. The hard, soft, and jet 
functions and the interactions of these modes are familiar objects  
in SCET. In addition, the effective theory after integrating out hard modes 
includes a resonant mode that describes squarks with off-shellness 
of order $\lambda$, a situation that is described by 
unstable-particle effective theory \cite{Beneke:2003xh}.
The soft-collinear physics is reminiscent of event shapes 
in $e^+ e^-$ annihilation in the phase-space region of two-jet 
final states. However, in the edge region of the cascade decay the two jets 
do not emanate from a point-like vertex, but from two points, 
the production and the decay vertices of the long-lived resonance. 
As a consequence the soft physics is much more complicated.

\begin{figure}[t]
  \centering
  \includegraphics[width=.9\textwidth]{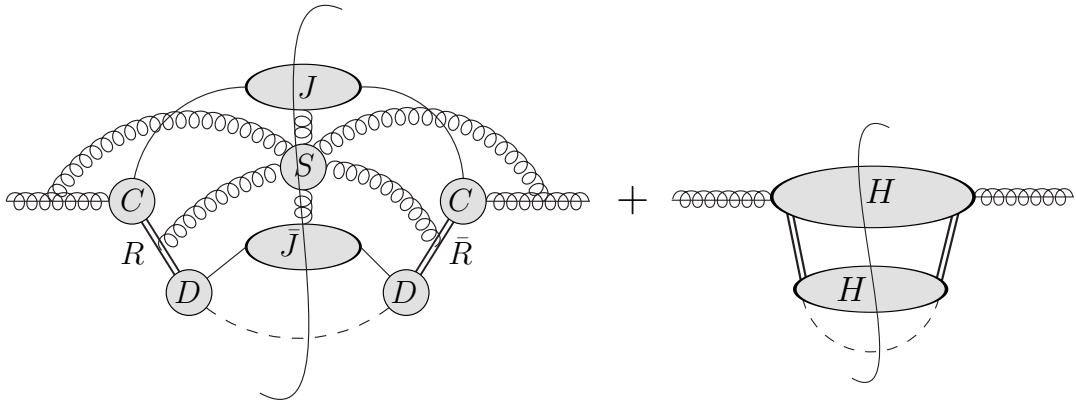}
\caption{Graphical representation of the factorization formula for the 
hadronic invariant mass distribution in the edge region. 
\label{fig:factformula}}
\end{figure}

We can therefore write down a factorization formula for the hadronic 
mass distribution of the form 
\begin{equation}
\frac{d\Gamma}{dM_h^2} = |C|^2 \cdot |D|^2 \cdot
J \otimes \bar J \otimes \left[R \otimes \bar R \otimes S\right] 
+ HR\,,
\label{eq:factformula}
\end{equation}
which is valid in the vicinity of $\medge^2$, at leading power in 
$\gasq/\msq$, and graphically presented in Figure~\ref{fig:factformula}. 
The first two factors on the right-hand side of this equation consist of the 
square of two hard functions, one ($C$) containing the hard virtual 
correction to the on-shell gluino decay $\sgl\to \sq+\bar q$, the other 
($D$) to the on-shell squark decay $\sq\to q+\neutralino$. The hard functions 
multiply the (anti)quark jet functions $J$, $\bar J$, which contain the 
collinear ($J$) and anti-collinear ($\bar J$) modes. These are 
convoluted with a soft function that consists of the resonant squark 
propagators $R$, $\bar{R}$ in unstable-particle effective theory, and 
a vacuum matrix element $S$ of soft Wilson lines factored 
off the jets, the gluino, and the resonances. Due to the spatial separation of 
the two hard decay vertices, the soft function is a highly non-local object 
in position space, which accounts for the distribution 
of soft momentum between the various factors and for the shape of the 
resonance.

The first term in (\ref{eq:factformula}) would be all there is, if 
requiring $M_h^2\approx \medge^2$ {\em always} forced the hadronic final 
state to consist of two back-to-back jets. However, hard 
(i.e. non-collinear) real emission is also possible. While the 
interference of hard emission amplitudes {\em between} the two decay stages 
is power-suppressed, since at least one squark propagator is then thrown 
off resonance, interference {\em within} the two decay stages separately 
can leave the squarks on-shell. In terms of Feynman diagrams 
the hard parton final state can therefore be described as a cluster of 
partons with invariant mass $p_J^2$ emerging from the gluino decay 
vertex and another cluster with mass $p_{\bar J}^2$ from the squark decay 
vertex, replacing the antiquark and quark in the tree diagram, 
respectively, such that the total invariant mass is near $\medge^2$. 
Note that $p_J^2$ and $p_{\bar J}^2$ are now generically of ${\cal O}(1)$  
while only the phase-space region when both are ${\cal O}(\lambda)$ is 
included in the first term on the right-hand side of (\ref{eq:factformula}). 
The additional hard-real contribution is denoted by 
$HR$ in this equation and must be added by explicit matching. In practice, 
this amounts to the calculation of hard real radiation to the 
separate decay stages in dimensional regularization, setting the external 
squark line on-shell and $M_h^2$ to $M_{\rm edge}^2$. These 
simplifications automatically avoid double counting with the first 
term of (\ref{eq:factformula}) and correspond to the direct computation 
of the hard region according to the method-of-regions strategy 
\cite{Beneke:1997zp}.

An intuitive understanding of hard real radiation is obtained from looking at 
the maximal value of $M_h^2$ for on-shell squarks in the presence of hard 
radiation. If this value is larger than $\medge^2$, the previous edge value 
lies in the bulk region of the hard radiative process.\footnote{This 
criterion also implies that at the level of one-gluon emission the hard gluon
 must be emitted from squark decay for light neutralinos 
$m_{\chiindex}<\msq^2/\msgl$, and from gluino decay for 
$m_{\chiindex}>\msq^2/\msgl$.} The latter therefore becomes insensitive to 
the width of the squark and can be treated like the bulk distribution at 
tree level. The $HR$ term is therefore simply a constant 
contribution to $d\Gamma/d M_h^2$ in the edge region. The structure of 
(\ref{eq:factformula}) is similar to resonant and 
non-resonant production in the factorization formula for the 
line-shape of a resonance or pair production near threshold 
in previous applications of unstable-particle effective theory (see the 
review \cite{Beneke:2015vfa}). However, here both terms appear at leading 
power due to the presence of a resonant bulk region at tree-level 
rather than a single resonant invariant mass or threshold energy. 
Hence the non-resonant contribution is replaced by the resonant, but 
width-insensitive and unsuppressed hard-real contribution. 

We shall provide a formal discussion of the factorization formula together 
with technical details in a separate paper \cite{longpaper}. 


\section{NLO invariant mass distribution}
\label{sec:NLO}

At leading order, the factorization formula (\ref{eq:factformula}) becomes 
trivial. Without any additional gluon, there is no hard radiation. The soft
and jet functions are unity. Defining the product of 
spin-averaged/spin-summed tree-level squared matrix elements 
$M^2(\hat{q}^2)=|M(\sgl\to\sq+\bar q)|^2 |M(\sq\to q+\neutralino)|^2$, 
the hard functions are given by $M^2(0)$, i.e.\ for 
vanishing off-shellness $\hat{q}^2=q^2-\msq^2=0$ of the squark. 
Since the on-shell kinematics is completely fixed at tree level, $M^2(0)$ is 
constant in phase space. The resonance factor $R$ is the
propagator with a constant width $\gasq$, so that
$R$ and $\bar R$ combine to a Breit-Wigner distribution. The integration 
of this distribution with respect to the off-shellness $\hat{q}^2$ of the 
squark is the only non-trivial phase-space integral. The measurement function 
for the hadronic mass introduces $\Theta$-functions
$\Theta\left(\hat{q}^2_\mathrm{max} - \hat{q}^2\right)$ and 
$\Theta\left(\hat{q}^2-\hat{q}^2_\mathrm{min}\right)$, which determine the 
integration range. Depending on the sign of 
$\chi=(\msq^4-\msgl^2 m_{\chiindex})/\msq^4$, either 
$\hat{q}^2_\mathrm{min}$ or $\hat{q}^2_\mathrm{max}$ is $\mathcal{O}(1)$, 
and the small off-shellness $\hat{q}^2$ can be neglected. Hence, the 
corresponding $\Theta$-function is always equal 
to one and can be omitted. The integration boundary in the other
$\Theta$-function can be expanded to leading order in $\lambda$.
Hence, the general result (\ref{eq:factformula}) simplifies to
\begin{equation}
\label{eq:LOformula}
\frac{d\Gamma_{\rm LO}}{d M_h^2} = 
\frac{M^2(0)}{256 \pi^3 \msgl^3} 
\int_{-\infty}^{\infty} d\hat{q}^2\, 
\frac{
\Theta\left(-\Delta - \hat{q}^2 \chi\right)}
{\hat{q}^4+\msq^2 \gasq^2} \, ,
\end{equation}
where  $\Delta$ is the $\mathcal{O}(\lambda)$ distance to the edge,  
i.e.\ $M_h^2 = \medge^2 + \Delta$. The $\Theta$-function results in a
universal tree-level shape of the edge distribution, since the dependence 
on the specific decay process appears only in the constant overall 
factor $M^2(0)$. 

We remark that at leading power in $\gasq/\msq$, the $\Theta$-function in 
(\ref{eq:LOformula}) is absent in the bulk region, since the second 
integration boundary is also $\mathcal{O}(1)$. Therefore, the differential 
width is constant in the bulk. In the tail region, on the other hand, 
the $\Theta$-functions make the differential width vanish.

\subsection{Next-to-leading power at tree level}
\label{sec:nlp}

Before turning to the calculation of the radiative corrections, we briefly 
discuss how the next term in the expansion in $\gasq/\msq$ of the tree-level 
distribution is computed in the edge region. At $\mathcal{O}(\lambda)$, the 
numerator $M^2(\hat{q}^2)$ in (\ref{eq:LOformula}) is not only needed for 
on-shell decays ($\hat{q}^2 = 0$) but one needs the next order in the 
Taylor-expansion of the off-shell matrix elements with respect to 
$\hat{q}^2$. Hence, one part of the resonant contribution is given by
\begin{equation}
\label{eq:NLOformulares}
\frac{d\Gamma_{\rm NLO}^{\rm res}}{d M_h^2} = 
\frac{dM^2(\hat{q}^2)/d\hat{q}^2|_{\hat{q}^2=0}}{256 \pi^3 \msgl^3} 
\int_{-\infty}^{\infty} d\hat{q}^2\, 
 \left(\frac{\hat{q}^2}{\mu^2}\right)^{\!\epsilon}\,  
\Theta\left(-\Delta - \hat{q}^2\chi\right) 
\,\, \frac{
 \hat{q}^{2}}
{\hat{q}^4+\msq^2 \gasq^2} \, ,
\end{equation}
where we have introduced the factor $\hat{q}^{2\epsilon}$ in order to make 
the integral well-defined. For $\chi>0$, the previous expression 
evaluates to 
\begin{equation}
\label{eq:NLOformularesexplicit}
\frac{d\Gamma_{\rm NLO}^{\rm res}}{d M_h^2} = 
\frac{dM^2(\hat{q}^2)/d\hat{q}^2|_{\hat{q}^2=0}}{256 \pi^3 \msgl^3} 
\left[ 
\frac{1}{\epsilon} + i \pi
+\frac{1}{2} \ln\left(
\frac{\msq^2 \gasq^2+\Delta^2/\chi^2}{\mu^4} \right)
\right] \, ,
\end{equation}
where the $1/\epsilon$ pole and the spurious imaginary part
are a consequence of factorizing the NLO contribution into a resonant and 
a non-resonant part, and of the choice of the regulating factor. Further, 
also the relevant integration boundary 
$\hat{q}^2_\mathrm{max}$ or $\hat{q}^2_\mathrm{min}$ discussed in the context 
of~(\ref{eq:LOformula}) receives an $\mathcal{O}(\lambda)$ correction, which 
can be taken into account by appropriately expanding the $\Theta$-function.
In addition, non-resonant contributions (with $|\hat{q}^2| \gg \msq\gasq$) 
start to contribute at NLO. Here, the Breit-Wigner propagator can be expanded 
in the small width, and the $\Theta$-function with respect 
to the small integration boundary. On the other hand, the second integration 
boundary can no longer be taken to infinity, and one also needs the full 
$\hat{q}^2$-dependence of the matrix elements. Hence, for $\chi>0$ one 
finds 
\begin{equation}
\label{eq:NLOformulanonres}
\frac{d\Gamma_{\rm NLO}^{\rm non-res}}{d M_h^2} = 
\frac{1}{256 \pi^3 \msgl^3} 
\int_{\hat{q}^2_{\rm min}}^{0} d\hat{q}^2\, 
 \left(\frac{\hat{q}^2}{\mu^2}\right)^{\!\epsilon}\, 
\,\, \frac{M^2(\hat{q}^2)}{\hat{q}^4} \, ,
\end{equation}
where we consistently applied the same regulating factor 
$\hat{q}^{2\epsilon}$ as above to render the integral well-defined, and 
$\hat{q}^2_{\rm min}$ is calculated for $M_h^2=\medge^2$. Since 
$M^2(\hat{q}^2)$ is polynomial in $\hat{q}^2$ in our case,
the integral can be easily computed. Combining the resonant and the
non-resonant contribution, the $1/\epsilon$ poles and spurious imaginary parts
contained in (\ref{eq:NLOformulares}) and (\ref{eq:NLOformulanonres}) 
cancel and the regulator can be set to zero.

In the tail and in the bulk regions, the non-resonant 
contribution is integrated from $\hat{q}^2_{\rm min}$ to 
$\hat{q}^2_{\rm max}$, where $\hat{q}^2_{\rm max/min}$ is determined as a 
function of $M_h^2$.  In the bulk, there is also a resonant contribution 
given by (\ref{eq:NLOformulares}) without the $\Theta$-function.
In all three regions, higher-order contributions in $\gasq/\msq$ are obtained 
in a straightforward way by expanding the relevant quantities (matrix 
elements, $\Theta$-functions, Breit-Wigner propagators) to the appropriate 
order.

For the specific SUSY process under consideration the interference between the 
diagrams in Figure~\ref{fig:cascade} starts to contribute at 
NLO in $\gasq/\msq$. There are two resonant contributions, where either the 
squark or the antisquark propagator is resonant and the other is off-shell, 
and quantities of $\mathcal{O}(\lambda)$ can be neglected in the remaining 
matrix element. There is also a non-resonant contribution, where $\hat{q}^2$ 
is considered to be large and the width can be neglected in both propagators.

\begin{figure}[t]
  \centering
  \includegraphics[angle=-90,width=0.48\textwidth]{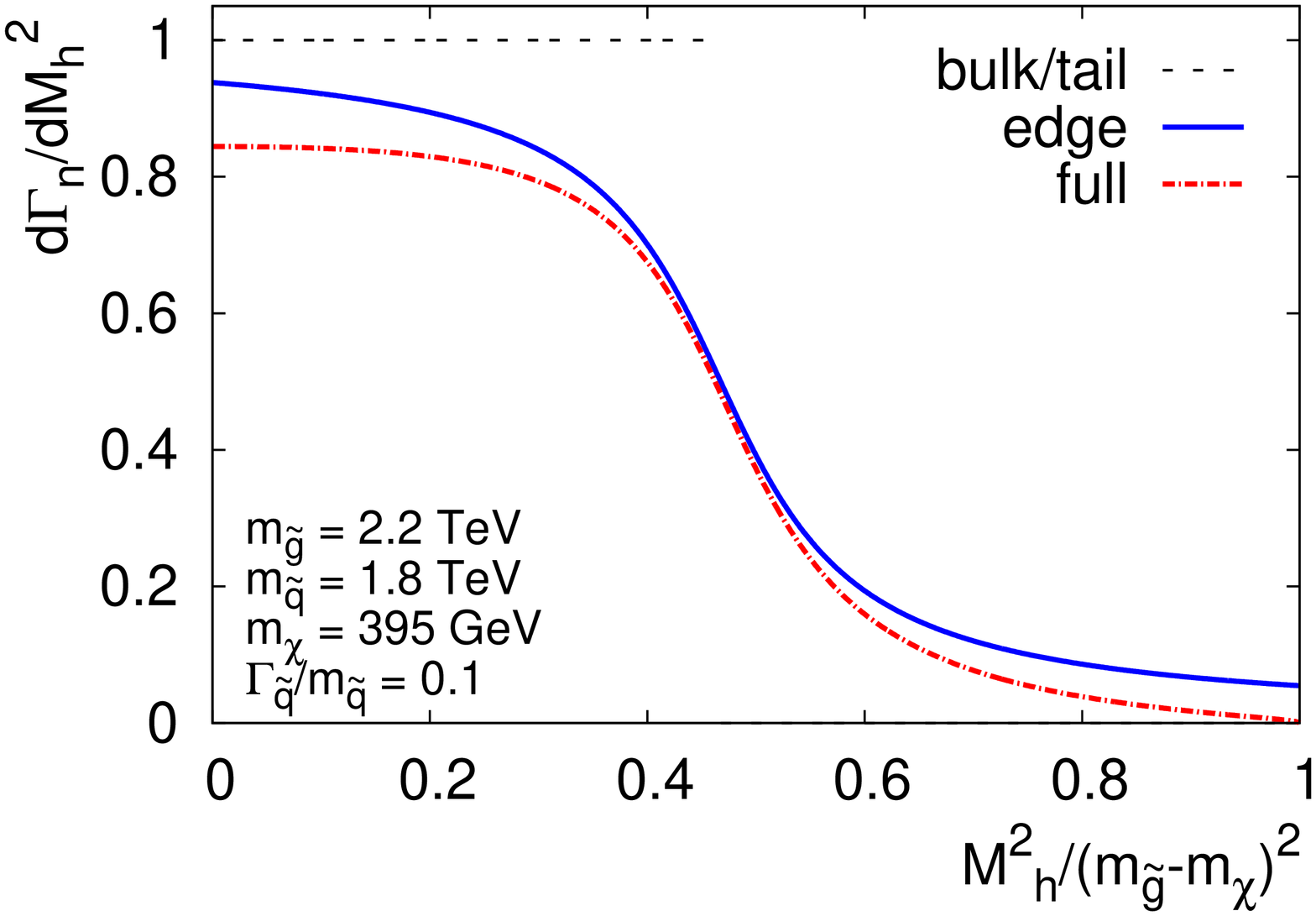}
  \includegraphics[angle=-90,width=0.48\textwidth]{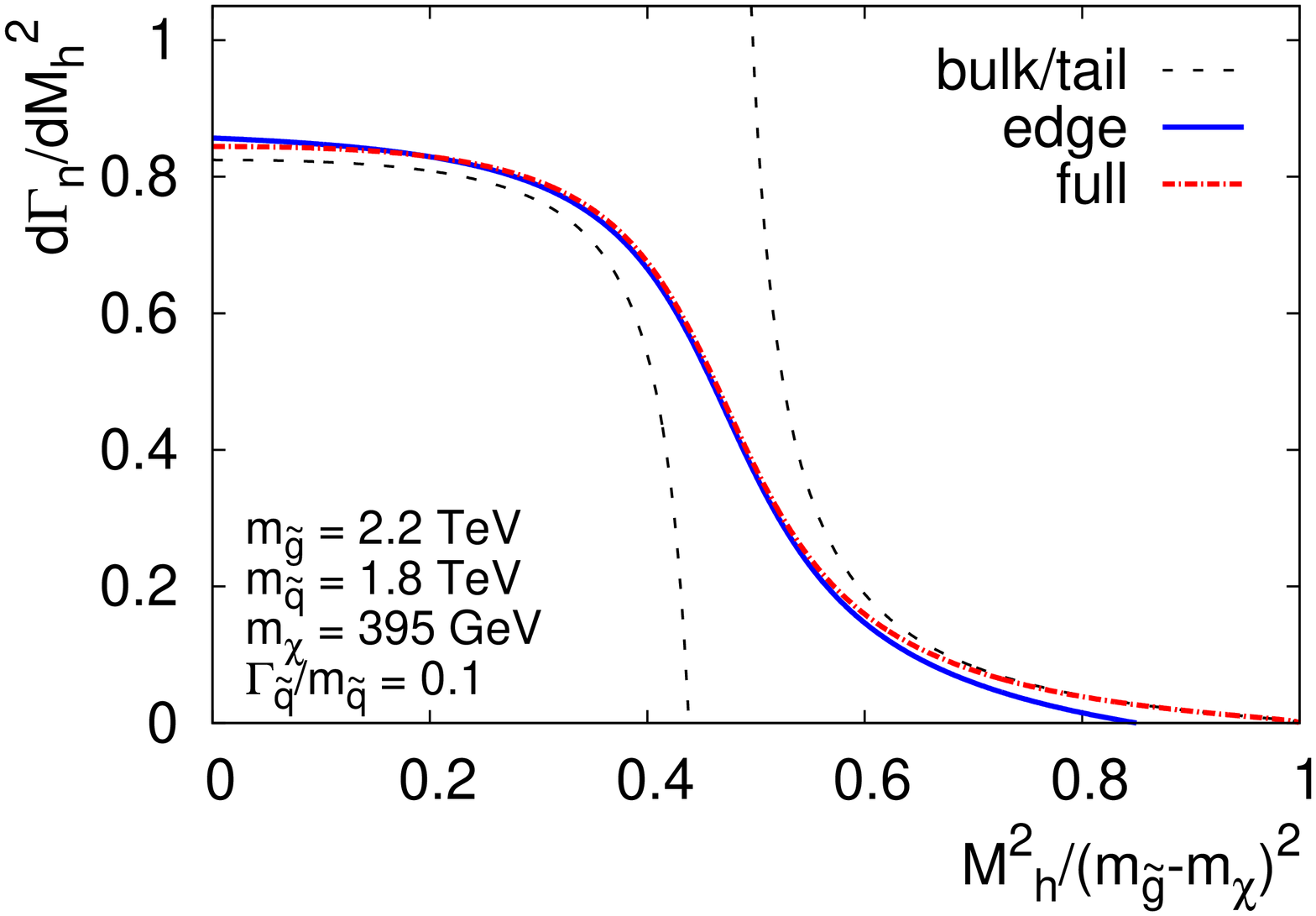}
  \includegraphics[angle=-90,width=0.48\textwidth]{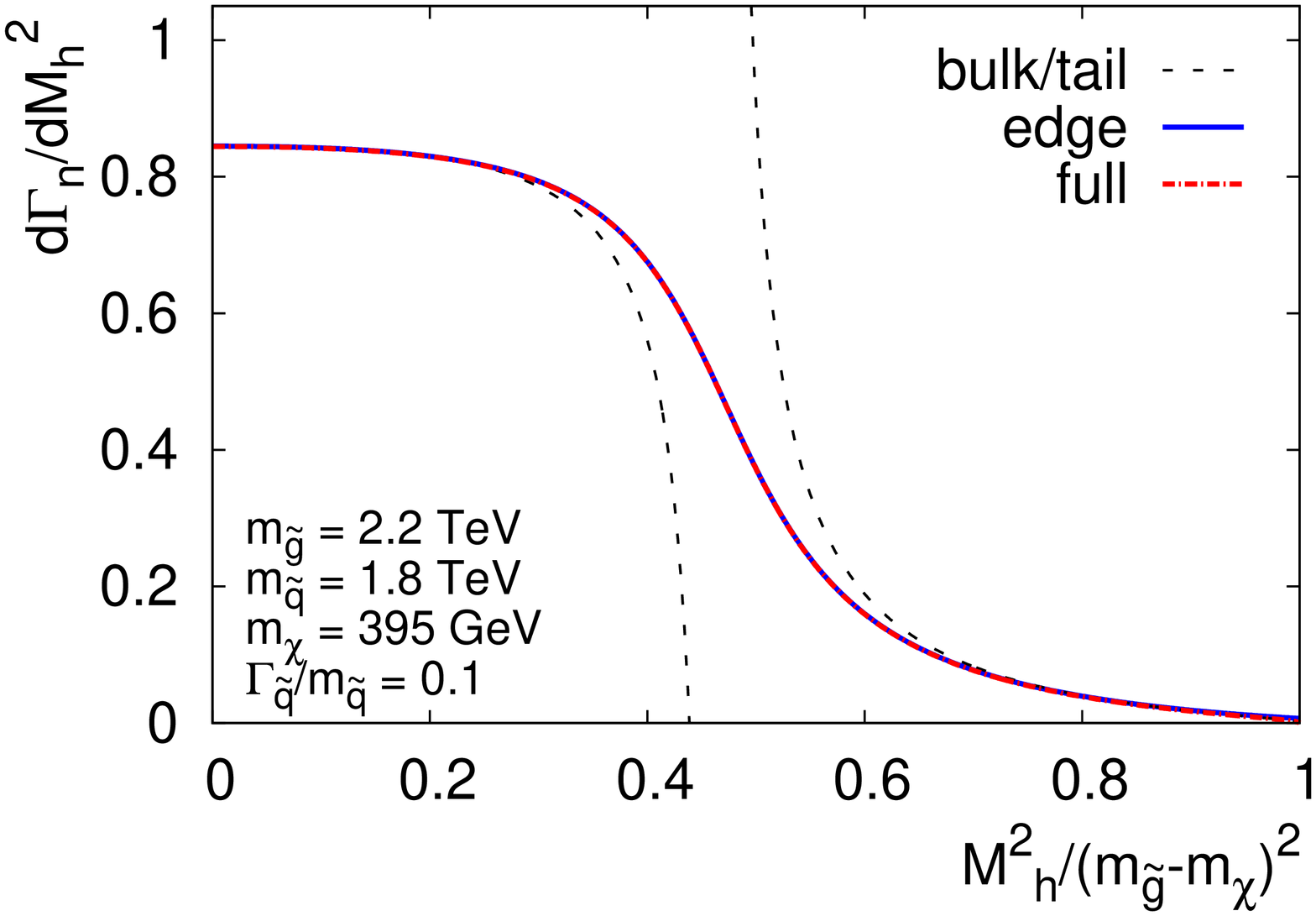}
\caption{Tree-level invariant mass spectrum for the SUSY benchmark point A 
with width $\gasq=\msq/10$ discussed below. We normalize the differential 
width to the constant LO result in the bulk region. 
The blue solid line refers to the edge distribution, 
the black dashed line to the tail and bulk distribution, and the red dot-dashed
line to the exact tree-level 
result in the fixed-width scheme. The interference
contribution is not included. From left to right to bottom the 
leading power (left), next-to-leading power (right),
and next-to-next-to-leading power (bottom) terms are successively 
included.  
\label{fig:treeedgewithNNLP}}
\end{figure}

In Figure~\ref{fig:treeedgewithNNLP}, we show numerical results for the 
tree-level hadronic invariant mass distribution at leading, next-to-leading, 
and next-to-next-to-leading power in $\gasq/\msq$ for the SUSY benchmark 
point A discussed below, and compare them to the exact tree-level result. 
Rapid convergence upon including higher powers in $\gasq/\msq$ 
can be observed.

\subsection{Radiative correction}
\label{sec:Radiativecorrection}

At next-to-leading order in the strong coupling $\alpha_s$, an additional 
gluon line is attached in all possible ways to the square of the tree 
diagrams shown in Figure~\ref{fig:cascade}. Hard virtual contributions 
amount to the evaluation of virtual corrections to each of the two two-body 
decays with an on-shell squark. They are also part of a standard narrow-width 
calculation~\cite{Beenakker:1996dw} and we do not discuss them further here. 
In (\ref{eq:factformula}) they give the NLO corrections to the hard 
functions $C$ and $D$, or equivalently the squared matrix elements $M^2(0)$ 
in (\ref{eq:LOformula}). In addition, there are soft, collinear and hard-real 
corrections which correspond to the expansion of the soft function and the 
jet functions, and to the evaluation of the hard real contribution in 
(\ref{eq:factformula}), respectively. They are discussed below.

All the individual pieces are in general separately divergent. 
When a soft gluon momentum $r$ flows through a squark propagator with 
momentum $k$ it is convenient to separate the UV divergent piece 
by adding and subtracting a term, such that the UV divergent term 
does not depend on the width and on $\hat{q}^2 = k^2-\msq^2$: 
\begin{equation}
\frac{1}{(k+r)^2-\msq^2+i \msq^2 \gasq^2} \stackrel{\mathcal{O}(\lambda)}{=}
\frac{1}{\hat{q}^2+2 \hat{k}\cdot r+i \msq^2 \gasq^2} =
\underset{\rm soft \,\, UV}{\underbrace{\frac{1}
{2 \hat{k}_{\phantom{\msq}}\!\!\!\!\!\!\cdot r}}} + 
\underset{\rm soft \,\, remainder}{\underbrace{
\frac{1}{\hat{q}^2+2 \hat{k}\cdot r+i \msq^2 \gasq^2} -
\frac{1}{2 \hat{k}\cdot r}}} \, , 
\end{equation}
where in the first step only terms of $\mathcal{O}(\lambda)$ are kept
in the denominator. Hence, one can neglect the gluon momentum in the 
off-shellness $\hat{q}^2$, and set the squark momentum to its 
value $\hat{k}$ in the on-shell $1\to 2$ gluino decay. 
In the following we separate the soft contributions into a term 
denoted ``soft UV'', which is simple and contains the UV divergence, 
and a finite ``soft remainder'' according to the above equation.
If no soft gluon momentum flows through a squark propagator, the
complete diagram is included in the soft UV contribution. 

For collinear gluon exchange the complete matrix elements in the
collinear approximation factor into the tree-level result and the
appropriate splitting function. After integration, 
the collinear contributions correspond to the convolution of the 
jet function in (\ref{eq:factformula}) with the resonant squark propagator.

For soft and collinear gluon exchange the $\hat{q}^2$-integral in 
(\ref{eq:LOformula}) has to be supplemented by a convolution with the gluon 
momentum. The virtual soft UV and collinear contributions are scaleless. 
For the real-emission  soft UV and collinear contributions, the
convolution can be cast into the standard form 
\begin{equation}
B(\Delta,n,c) =\left(\frac{\mu}{\gasq}\right)^{\!n \epsilon}
\int_0^{\infty}dy\,y^{-1-n \epsilon}
\int_{-\infty}^{\infty}d\hat{q}^2 \, 
\frac{\Theta(-\Delta-c\, y\,\gasq \msq \, \chi-\hat{q}^2\chi)}
{\hat{q}^4+\msq^2 \gasq^2} \, ,
\end{equation}
where $\mu$ is the scale from dimensional regularization 
and $y$ is related to the small components of the soft or collinear gluon
momenta. Since $y$ enters the $\Theta$-function at NLO for real
gluon emission, the integral is not scaleless. In the bulk and in the tail 
regions, the argument of the $\Theta$-function is $\mathcal{O}(1)$ and $y$
must be neglected at leading power. Hence, in these regions there are no 
soft and no collinear contributions, as expected. The integral is given by 
\begin{eqnarray}
&&B(\Delta,n,c)
= -\frac{{\rm sgn}(\chi)}{\msq \gasq}\,
\Gamma(n \epsilon) \Gamma(-n\epsilon)
\, {\rm Im} \left[x^{-n\epsilon}\right] 
\nonumber \\
\label{eq:Bexpansion}
&& \hspace*{0.5cm} =  - \frac{{\rm sgn}(\chi)}{\msq \gasq} \,
{\rm arg}(x) \left(
\frac{1}{n\epsilon}- \ln |x|
- \frac{n \epsilon}{6} \left({\rm arg}^2(x) -\pi^2 - 3\ln^2|x|
\right)\right) + \mathcal{O}(\epsilon^2)
\,,
\qquad
\end{eqnarray}
where $x=\frac{\gasq}{\mu}\frac{\Delta/\chi+i\msq
\gasq}{c \, \msq \gasq}$.
The combined soft UV and collinear contribution to the differential width 
reads
\begin{eqnarray}
\label{eq:softUVonalcollinearformula}
\frac{d\Gamma_{\rm soft \,UV+coll}}{d M_h^2} = 
\frac{M^2(0)}{256 \pi^3 \msgl^3} 
\frac{ \alpha_s}{\pi} \,e^{\gamma_E\epsilon}\!\!\!
&& \left[
B(\Delta,2,c_i) A_i^{\rm soft \,UV} + B(\Delta,2,-c_f) A_f^{\rm soft \,UV} 
\right. 
\nonumber\\
&& \left. 
+ B(\Delta,1,c_i) A_i^{\rm coll} + B(\Delta,1,-c_f) A_f^{\rm coll} \right]\,,
\end{eqnarray}
where  $c_i=\msq/\msgl$, $c_f=\msgl/\msq$, and
\begin{align}
\notag
A_i^{\rm soft \,UV} = & \,
 C_{\A\X} \, \Gamma(\epsilon) - C_{\A\A} \, \Gamma(1+\epsilon)
- C_{\A\R} \, \Gamma(\epsilon) \frac{\msgl^2+\msq^2}{\msgl^2-\msq^2}\,
\left[1-\left(\frac{\msgl^2}{\msq^2}\right)^{\!\epsilon}\,\right]\\
& \, \label{eq:AisoftUV}
+C_{\R\X}\, \Gamma(\epsilon) \left(\frac{\msgl^2}{\msq^2}\right)^{\!\epsilon} 
- C_{\R\R} \, \Gamma(1+\epsilon)
\left(\frac{\msgl^2}{\msq^2}\right)^{\!\epsilon} \, ,\\
A_f^{\rm soft \,UV} = & \,\label{eq:AfsoftUV}
 C_{\R\Y} \, \Gamma(\epsilon) \left(\frac{\msq^2}{\msgl^2}\right)^{\!\epsilon} 
- C_{\R\R}\, \Gamma(1+\epsilon)
	\left(\frac{\msq^2}{\msgl^2}\right)^{\!\epsilon} \, ,\\
A_i^{\rm coll} = & 
\label{eq:Aicoll}
\, C_F \left(\frac{\mu}{\msgl}\right)^{\!\epsilon} 
\left(\frac{\msgl^2-\msq^2}{\msgl^2}\right)^{\!-\epsilon} 
\frac{\Gamma(2-\epsilon)}{\Gamma(1-\epsilon)}\left( \frac{\Gamma(-\epsilon)}
{\Gamma(2-2\epsilon)} +
\frac{\Gamma(2-\epsilon)}{2\Gamma(3-2\epsilon)}\right)\, ,\\
A_f^{\rm coll} = & 
\label{eq:Afcoll}
\, C_F \left(\frac{\mu}{\msgl}\right)^{\!\epsilon} 
\left(\frac{\msq^2-m_{\chiindex}}{\msq^2}\right)^{\!-\epsilon} 
\frac{\Gamma(2-\epsilon)}{\Gamma(1-\epsilon)}\left( \frac{\Gamma(-\epsilon)}
{\Gamma(2-2\epsilon)} +
\frac{\Gamma(2-\epsilon)}{2\Gamma(3-2\epsilon)}\right)\, .
\end{align}
Here the $C_{ij}$ correspond to the colour factors of the diagram with the
gluon attached to $i$ and $j$. To be specific $C_{\A\A}=N_c$, 
$C_{\A\X}=C_{\A\R}=C_{\A\Y}=N_c/2$,
$C_{\X\X}=C_{\Y\Y}=C_{\R\Y}=C_{\R\R}=C_F$,
$C_{\R\X}=C_{\X\Y}=C_F-N_c/2$ and in turn $C_F=4/3$, $N_c=3$. The different
factors for the individual soft UV pieces are due to the angular integrals 
over the soft UV propagators. In the collinear case the $\Gamma$-functions
are due to the integral over the collinear momentum fraction of the 
emitted gluon.

The coefficients in (\ref{eq:AisoftUV})-(\ref{eq:Afcoll}) contain 
$1/\epsilon$ poles which combine with the pole in (\ref{eq:Bexpansion})
to cancel the poles of the hard virtual and hard real contributions.
The above results are obtained with a  
($4-2\epsilon$)-dimensional phase space for the gluon but with a 
4-dimensional phase-space for the particles present at tree-level. We have
verified that the final result, including the hard contributions 
calculated using the same convention, agrees with the result in
conventional dimensional regularization.

The leading logarithmically enhanced corrections of the form 
$\ln^n(\msq/\gasq)$ ($n \le 2$) in the full result can be extracted from 
(\ref{eq:softUVonalcollinearformula}) alone using the $\epsilon$-expansion
in (\ref{eq:Bexpansion}). If $\mu$ is chosen $\mathcal{O}(\msq)$, e.g. 
$\mu=\msq$, all large logarithms are contained in the soft UV+collinear 
part, since then the hard pieces depend only on $\mathcal{O}(1)$ ratios of 
dimensionful parameters. The soft remainder discussed below also does not 
contain large logarithms since it is finite, $\mu$-independent, and 
homogeneous in the soft scale. 

Due to the appearance of the width and $\hat{q}^2$ in the squark propagator, 
the virtual soft-remainder contributions are not scaleless. We evaluate the 
virtual diagrams by taking residues to convert them into phase-space diagrams. 
They then combine with 
the soft remainder from the real diagrams such that most infrared  
divergences cancel. Some diagrams,  however, show a purely imaginary pole
in the remainder which vanishes after adding the complex conjugate diagram.

The finite soft remainder contribution can be 
expressed in terms of a one-dimensional integral representation based on a 
single standard integral, i.e.\
\begin{eqnarray}
\frac{d\Gamma_{\rm soft\,\, remainder}}{d M_h^2} & = &
\,\frac{M^2(0)}{256 \pi^3 \msgl^3} 
\frac{\alpha_s}{4 \pi} \,\, {\rm sgn(\chi)} \int_{-1}^{1} {\rm d} x
\nonumber\\ 
&& \hspace*{-2cm} \times
\left[
\left( 
\frac{C_{\A\R}}{\alpha} \, \frac{\msgl^2+\msq^2}{2\msgl^2} 
+ \frac{C_{\R\X}-C_{\X\Y}}{(1-x)\,\alpha} 
\right)
F\!\left(\Delta,\frac{(1-x)\,c_i\,\msq}{\alpha\, \msgl}\right) \right. 
\nonumber\\ 
&&  \hspace*{-2cm} \left. 
+ \left(
\frac{C_{\R\Y}-C_{\X\Y}}{(1+x)\, \alpha} \, 
\frac{\msq^2}{\msgl^2}-\frac{C_{\A\Y}}{1+x} -
\frac{C_{\R\R}}{\alpha^2}\, \frac{\msq^2}{\msgl^2} \right) 
F\!\left(\Delta,\frac{(1+x)\,c_f\,\msq}{\alpha\, \msgl}\right) +{\rm c.c.}
\right],
\end{eqnarray}
where $\alpha=((1+x)\msgl^2+(1-x)\msq^2)/(2\msgl^2)$ and 
\begin{equation}
F(\Delta,c)= I_+(\Delta,c-2)-I_+(\Delta,-2)-I_-(\Delta,c)
\end{equation}
with 
\begin{equation}
I_\pm(\Delta,\beta) = \frac{1}{i\msq \gasq} 
\left[\frac{\pi^2}{6} - 
{\rm Li}_2\left(1+\beta\frac{i\, \msq \gasq}{-\Delta/\chi\pm i \, \msq \gasq} 
\right) \right] \, .
\qquad
\end{equation}
In particular, diagrams where the gluon connects the gluino or the antiquark 
to the quark from squark decay only consist of these finite contributions. 

Taking residues to evaluate the virtual loop diagrams, there are so-called 
particle-pole contributions (poles not due to the gluon propagator)
which need additional analytic regularization to render separate soft 
and Glauber regions well-defined. All particle pole contributions vanish 
when properly regularized.

The hard real gluon emission contributes to the second term in 
(\ref{eq:factformula}). For hard gluon momenta, there is no non-trivial
convolution between the squark and the gluon momentum, since $\hat{q}^2$ 
can be neglected in the argument of the measurement function,
which is $\mathcal{O}(1)$ for a hard real gluon. 
For the same reason $\Delta$ can be neglected, and the resulting  
$d\Gamma_{\rm hard\,real}/dM_h^2$ is a constant in the
edge region. To compute $HR$, one has to compute the phase-space integral
over the real-emission matrix elements for on-shell squarks as in a standard
narrow-width calculation. Only the gluino (large $m_{\chiindex}$, $\chi<0$) or 
the squark decay (small $m_{\chiindex}$, $\chi>0$)
contribute depending on the sign of $\chi$, since the measurement function
restricts the possible values for the quark-gluon and antiquark-gluon
invariant masses for $M_h^2 = \medge^2$. It is convenient to use the  
following subtraction procedure. As usual the differential width is written 
as a phase-space integral over the squared matrix element involving the 
measurement function. We add and subtract the full squared matrix element 
divided by $\medge^2$ without applying the
measurement function, i.e.
\begin{equation}
\label{eq:hardsubtraction}
|M|^2 \delta(\medge^2-M_h^2) = 
\left(|M|^2 \delta(\medge^2-M_h^2) - |M|^2/\medge^2\right) +
|M|^2/\medge^2 \, .
\end{equation}
We first perform the phase-space integral over the angle between the quark and 
the antiquark. While the subtraction term is independent of this angle and can
be trivially integrated, using the $\delta$-function to perform the angle 
integral leads to non-trivial phase-space boundaries and a phase-space 
dependent factor multiplying the matrix element. In the soft-collinear 
phase-space region, which is always contained in the integration range, this 
factor tends to $1/\medge^2$. Hence,
the remaining phase-space integration over the subtracted piece in 
parenthesis in (\ref{eq:hardsubtraction}) is 
finite in four dimensions and can be
easily computed (we use a one-dimensional integral representation for our
numerical results). The remaining term in (\ref{eq:hardsubtraction}) is 
proportional to the real-emission contribution to the total width upon 
integration. It has to be calculated
using dimensional regularization but it is known from inclusive 
narrow-width calculations.

The hard virtual, the hard real, and the soft UV and collinear 
contributions in (\ref{eq:softUVonalcollinearformula}) individually include 
poles in $1/\epsilon$. 
In contrast to an inclusive calculation, the poles in the hard virtual and 
hard real pieces do not cancel at the edge because of the non-trivial 
$\hat{q}^2$ integral (\ref{eq:LOformula}), which multiplies the hard virtual 
correction. Together with the soft and collinear contributions in 
(\ref{eq:softUVonalcollinearformula}), the differential width is,
of course, finite (diagram by diagram if the collinear contributions are
split accordingly). The sum of the hard virtual and hard real corrections for 
the inclusive calculation agrees with the results in~\cite{Beenakker:1996dw}.

\begin{figure}[p]
  \centering  
  \includegraphics[angle=-90,width=.51\textwidth]
{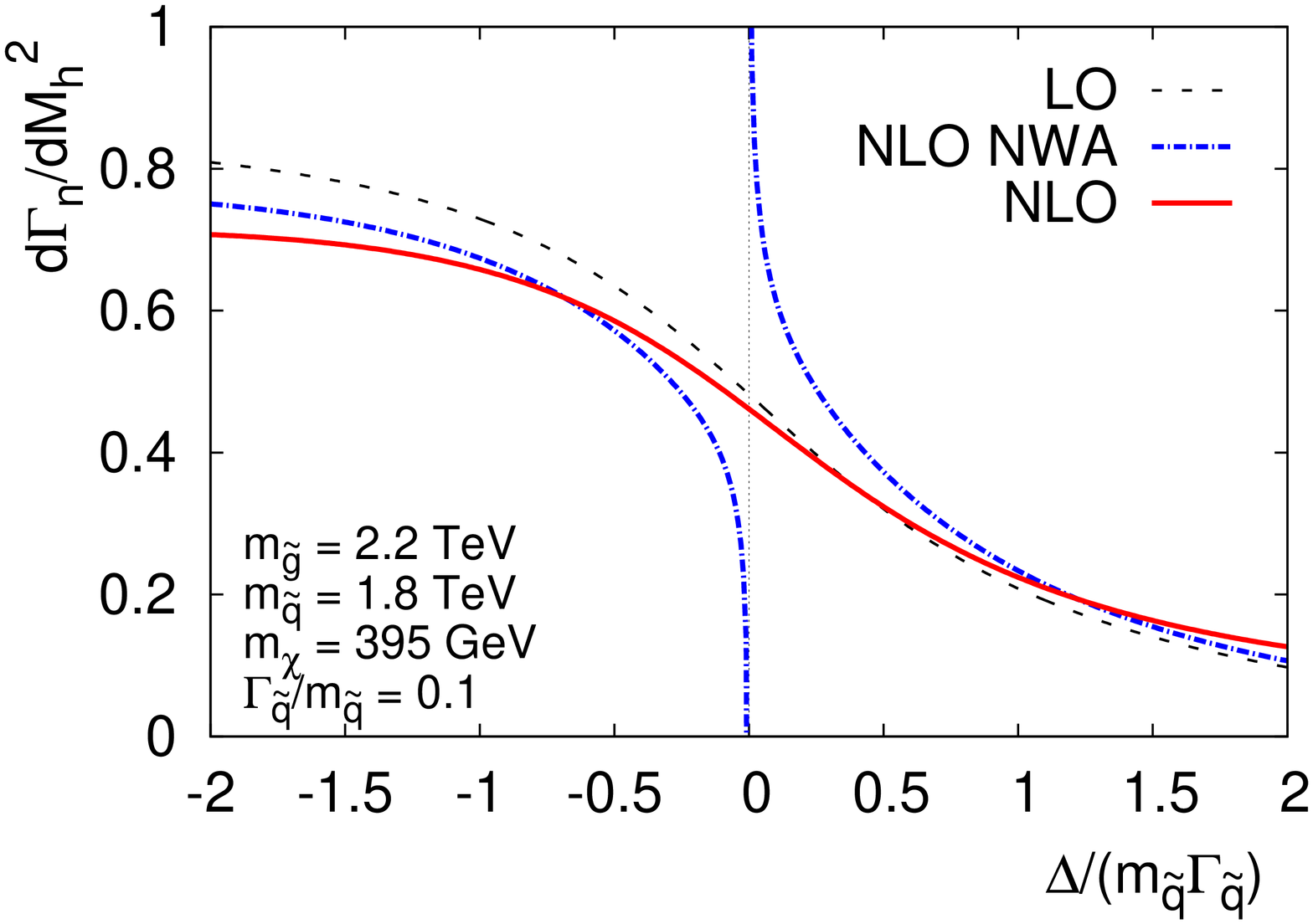}\\
  \includegraphics[angle=-90,width=.51\textwidth]
{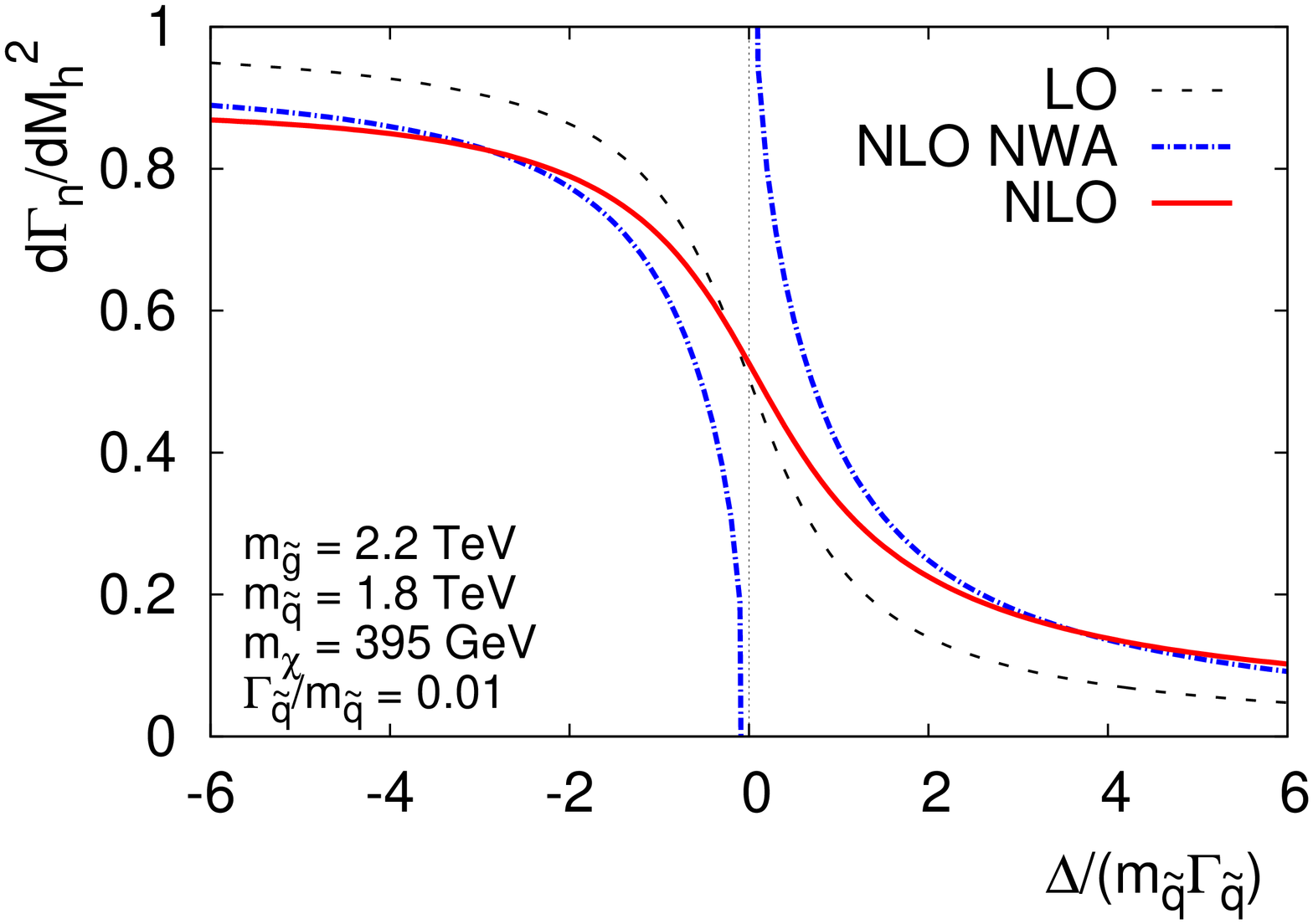}\\
  \includegraphics[angle=-90,width=.51\textwidth]
{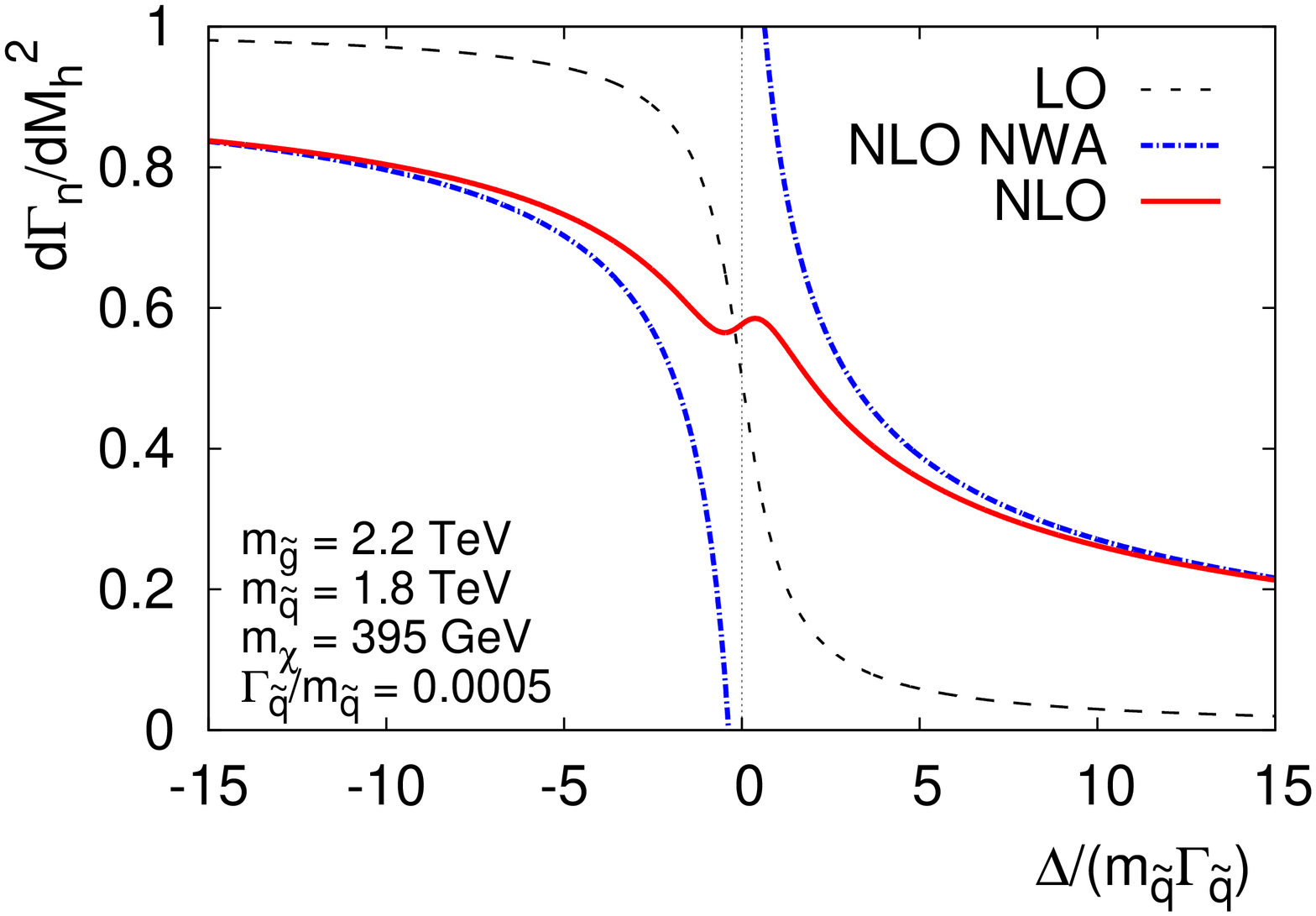}\\
\caption{\label{fig:NLO_numerical_result_A} 
Differential width as a
function of the distance to the edge for the benchmark
scenario A: $m_{\sgl}=2.2$ TeV, $\msq=1.8$ TeV, $m_{\chiindex} = 395$
GeV, $\chi>0$. As in Figure~\ref{fig:treeedgewithNNLP}, we normalize 
the differential width to the constant LO bulk region result in 
the narrow-width approximation. Shown are (black dashed) the full LO result 
in $\alpha_s$ (including all power corrections in $\gasq/\msq$), 
(red solid) the NLO QCD corrections added, which is our main result, 
and (blue dot-dashed) the full LO result plus the NLO QCD 
corrections in the narrow-width approximation in the bulk and in the 
tail region. From top to bottom the plots refer to 
three different choices of the width:
$\gasq/\msq=0.1$ (top), $\gasq/\msq=0.01$ (middle),
$\gasq/\msq=0.0005$ (bottom).}
\end{figure}

The choice of SUSY benchmark points for our numerical analysis is based
on the exclusion limits in terms of simplified models
provided by ATLAS~\cite{Aad:2015iea}. We analyzed two points which are 
not excluded, one for $\chi>0$ and one $\chi<0$, with parameters
\begin{itemize}
\item benchmark A: $\msgl =2.2$ TeV, $\msq=1.8$ TeV, $m_{\chiindex} = 395$
GeV, $\chi>0$
\item benchmark B: $\msgl=2.2$ TeV, $\msq=1.0$ TeV, $m_{\chiindex} = 695$
GeV, $\chi<0$
\end{itemize}
Note that all results depend only on the ratios of the masses and the 
squark width. We use the squark width as a free parameter to investigate the
edge behaviour for different values of $\gasq/\msq$. For the renormalization 
scale we use $\mu=\msgl$ with $\alpha_s(\mu)=0.0799$. In 
Figure~\ref{fig:NLO_numerical_result_A} we show the result for benchmark A. 
The scenario B exhibits similar features and is therefore 
not shown. The figures display the full LO result in $\alpha_s$ 
(including all power corrections) for the hadronic invariant mass 
distribution $d\Gamma_{\rm LO}/d M_h^2$ (black dashed). 
To this we add the NLO QCD corrections at leading power  $\gasq/\msq$ in 
the edge region (red solid), which is our main result, and the NLO QCD 
corrections in the narrow-width approximation in the bulk and in the 
tail region, which diverge at the edge, for comparison (blue dot-dashed).
The edge result gives a valid description where the 
narrow-width approximation at NLO in $\alpha_s$ 
clearly fails. At the edge, power corrections $\mathcal{O}(\gasq/\msq)$
are missing. Going away from the edge, further power corrections 
of $\mathcal{O}(\Delta/\msq^2)$ arise, eventually become dominant, and 
destroy the validity of the edge description. For large width (upper plot in 
Figure \ref{fig:NLO_numerical_result_A}), missing power corrections
are sizeable, as can be expected, such that there is no overlap region 
where the bulk/tail and edge results properly match.
For medium width (middle plots in Figure \ref{fig:NLO_numerical_result_A}), 
the results for the three regions agree reasonably well if $\Delta$ is a 
few times $\msq\gasq$. This is where the different approximations should 
be matched, since power corrections in the edge of the form $\Delta/\msq^2$ 
and logarithmically enhanced terms in the bulk/tail 
(see (\ref{eq:bulktaillogs})) are both subdominant. The matching of the 
bulk/tail and the edge results improves with decreasing width as can be seen 
by comparing the upper and the lower plots in 
Figure~\ref{fig:NLO_numerical_result_A}. However, for decreasing width, the 
logarithms of $\msq/\gasq$ in the edge result increase and resummation of 
these logarithms becomes mandatory for an extremely small width. The onset of 
the unphysical behaviour of the unresummed edge result can be seen close to 
the edge in the lower plot of Figure \ref{fig:NLO_numerical_result_A}.

\section{Conclusion}
\label{sec:conclusion}

Kinematic edges of cascade decays of new particles produced in high-energy 
collisions may provide important constraints on the particle masses. 
Depending on the experimental resolution an accurate treatment of 
finite-width and higher-order radiative effects is required. In this work we 
performed a next-to-leading order calculation in the two small 
quantities $\alpha_s$ and $\gasq/\msq$ for the hadronic invariant mass 
distribution in the vicinity of the kinematic edge of the gluino cascade decay 
$\tilde{g}\to q\bar q \neutralino$ through a squark resonance, 
based on a systematic expansion in $\gasq/\msq$.

At NLO it is of course technically possible to perform a standard 
one-loop computation in the complex mass scheme, as was done for the 
electromagnetic correction to the decay $\tilde{\chi}_2^0\to \ell\bar\ell 
\tilde{\chi}_1^0$ through a slepton resonance \cite{Drees:2006um}. 
The approach discussed here is nevertheless interesting, since 
the separation into hard, collinear and soft contributions does not only 
simplify the calculation, but also elucidates the process-dependent 
and universal features of distributions in the edge region. We then 
find that these are described in terms of on-shell decay matrix 
elements, universal jet functions and a soft function that depends 
only on the resonance propagator and soft Wilson lines, one for 
each coloured particle involved.

For very narrow resonances the perturbative approximation breaks down 
due to large width logarithms, a situation that becomes relevant 
only for exquisite experimental resolution. The factorization structure 
discussed here makes it clear that these logarithms can be summed 
with the help of renormalization group equations for the hard and 
jet functions. We hope to return to this in a future publication.

\subsubsection*{Acknowledgements}

The work of M.B. has been supported in part by the Bundesministerium f\"ur 
Bildung und Forschung (BMBF) under project no.~05H15W0CAA. L.J. was 
partially supported by the DFG contract STU 615/1-1, and M.U. is supported 
by a Royal Society Dorothy Hodgkin Research Fellowship.

\providecommand{\href}[2]{#2}\begingroup\raggedright\endgroup


\end{document}